\documentclass[twocolumn,showpacs,groupedaddress]{revtex4}
\usepackage{graphicx}
\usepackage{dcolumn}
\usepackage{bm}
\usepackage{epsfig} 
\usepackage{rotate}
\newcommand{\be}{\begin{equation}}
\newcommand{\ee}{\end{equation}}
\newcommand{\ba}{\begin{eqnarray}}
\newcommand{\ea}{\end{eqnarray}}
\newcommand{\etal}{{\em et al.}}
\usepackage{epstopdf}
\input{tcilatex}
\begin{document}
 \author{R. E. Warner}
 \affiliation{Oberlin College, Oberlin, Ohio 44074}
\author{F. Carstoiu,}
\affiliation{Cyclotron Institute, Texas A \& M University, College Station, Texas 77843}
\affiliation{IFIN-HH, 76900 Bucharest-Magurele, Romania}
\author{J. A. Brown}
\affiliation{Wabash College, Crawfordsville, Indiana  47933}
\author{F. D. Becchetti}
\affiliation{University of Michigan, Ann Arbor, Michigan 48109 }
\author{B. Davids}
\affiliation{TRIUMF, 4004 Wesbrook Mall, Vancouver, Canada, V6T 2A3 }
\author{A. Galonsky}
\affiliation{National Superconducting Cyclotron Laboratory, East Lansing, Michigan
48824 }
\author{M. Horoi}
\affiliation{Department of Physics, Central Michigan University, Mount Pleasant, Michigan
48859 } 
\author{J. J. Kolata}
\affiliation{University of Notre Dame, Notre Dame, Indiana 46556 }
\author{A. Nadasen}
\affiliation{University of Michigan, Dearborn, Michigan 48128 }
\author{D. A. Roberts}
\affiliation{University of Michigan, Ann Arbor, Michigan 48109 }
\author{R. M. Ronningen}
\affiliation{National Superconducting Cyclotron Laboratory, East Lansing, Michigan
48824 }
\author{C. Samanta}
\affiliation{Virginia Commonwealth University, Richmond, Virginia 23284}
\affiliation{Saha Institute of Nuclear Physics, 1/AF, Bidhannagar, Kolkata
700064}
\author{J. Schwarzenberg}
\affiliation{Institute of Nuclear Physics, University of Vienna, Waehringerstrasse 17, A-1090, Vienna, Austria}
\author{M. Steiner}
\affiliation{National Superconducting Cyclotron Laboratory, East Lansing, Michigan
48824 }
\author{K. Subotic}
\affiliation{Institute of Nuclear Sciences, VINCA, Belgrade 11001, Yugoslavia}

\title{Reaction and proton-removal cross sections of $^6$Li, $^7$Be, $^{10}$B, $^{9,10,11}$C, $^{12}$N, $^{13,15}$O and $^{17}$Ne on Si at
15 to 53 MeV/nucleon }

\date{\today}
\begin{abstract}
Excitation functions for total reaction cross sections, $\sigma_R$, were 
measured for the light, mainly proton-rich
nuclei $^6$Li, $^7$Be, $^{10}$B, $^{9,10,11}$C, $^{12}$N, $^{13,15}$O, and
$^{17}$Ne incident on a Si telescope at energies between 15 and 53
MeV/nucleon.  The telescope served as target, energy degrader and 
detector. Proton-removal cross sections, $\sigma_{2p}$ for $^{17}$Ne and
$\sigma_p$ for most of the other projectiles, were also measured.  The strong absorption model reproduces the $A$-dependence of $\sigma_R$, but not the detailed structure. Glauber multiple scattering theory and the JLM folding model provided improved descriptions of the measured $\sigma_R$ values. 
$rms$ radii, extracted from the measured $\sigma_R$
using the optical limit of Glauber theory, are in good agreement with those 
obtained from
high energy data. One-proton removal reactions are described using 
an extended
Glauber model, incorporating second order noneikonal corrections, 
realistic single particle densities, and  
spectroscopic factors from shell model calculations.
\end{abstract}

\maketitle

\section{Introduction}

The interesting properties of radioactive nuclei produced in the laboratory
include their lifetimes, sizes and distributions of nuclear matter, shell
structure, excited states, and decay modes.  
 Often, the one-particle separation energies are small, of the order of 1 MeV or less. These small separation energies lead to a wealth of phenomena including soft collective modes, exotic transition strengths between low-lying states, changes in shell structure, long-tailed density distributions, and, perhaps most dramatically, halo nuclei.
 In one-particle removal
reactions, the  parallel and transverse momentum distributions of the core-like fragments
are very narrow as compared with those of  normal nuclei, indicating increased nuclear size. The total reaction and breakup
cross sections are large, also reflecting the increased nuclear size. Spontaneous
$2p$ radioactivity has been observed recently near the proton drip
line\cite{pfu1,borc1}. The increased
density in the tails of the matter distribution generates a competition between the increased
refractive power of the real optical potential and the increased absorption due to the imaginary part,
leading to exotic shapes in heavy ion elastic scattering angular distributions.

Total reaction cross sections, $\sigma_R$, and proton-removal cross
sections, $\sigma_p$, contain complementary information about the size and
matter distributions of atomic nuclei. Various reactions at all impact
parameters contribute to $\sigma_R$, which therefore reflects mainly the
$rms$ nuclear radius.  Breakup, however, is a peripheral process, hence
$\sigma_p$ is sensitive mainly to the surface distribution. Therefore these
combined data are needed to better describe the matter distribution, and
the best test of theory is obtained by fitting them simultaneously.

Measurements of both $\sigma_R$ and $\sigma_p$ can be used to identify 
proton-halo nuclei. Among the light nuclei, $^8$B has been identified as a
proton-halo nucleus. The evidence includes its enhanced $\sigma_R$ at low
energies \cite{rew95},  as well as a narrow longitudinal momentum distribution of the $^7$Be core following proton removal. \cite{kelley,smed99,davids01}

We report $\sigma_R$ measurements for the stable and short-lived light
nuclei $^6$Li, $^7$Be, $^{10}$B, $^{9,10,11}$C, $^{12}$N, $^{13,15}$O, and
$^{17}$Ne incident on Si targets at energies ranging from 15 to 53
MeV/nucleon. Most of these nuclei are proton-rich; some are on the proton
drip line, and therefore may be proton-halo candidates. The measurements
were made by aiming these projectiles at a stack of thin Si elements which
served as both targets and detectors.  We also report $\sigma_p$ for most of the
projectiles, the exceptions being $^9$C, whose $\sigma_p$ we previously
reported \cite{rew04}, and  $^6$Li and $^{10}$C, which are unbound following
single proton removal. We report $\sigma_{2p}$ for the 2-proton-halo candidate $^{17}$Ne
\cite{kanungo}, which is also unbound after p-removal. A
preliminary account of this experiment appeared earlier \cite{rew98}.  Those
data are superseded by the present measurements since we have now more
effectively identified and rejected the unwanted components of the  beams used
in the experiment.

Proton and heavy ion radiotherapy is now utilized in the treatment of cancer
\cite{litzen}.  These beams produce nuclear
fragments, including short-lived isotopes such as $^{11}$C, and these can
contribute significant dosage.  Experimental $\sigma_R$ data are needed for
both stable and short-lived light ions to better quantify both the dosage
delivered by these fragments and their spatial distribution along the beam
path. 

We interpret our data using phenomenological strong absorption models, as
well as more elaborate models such as the optical limit of Glauber
multiple-scattering theory. Breakup reactions are described in  an extended 
Glauber model, incorporating second order
noneikonal corrections and shell model spectroscopic factors, as well as 
double folding optical potentials generated by complex, density-dependent and energy-dependent nucleon-nucleon (NN) effective interactions.

Section II of this paper describes the beams, target-detector system, and 
procedure utilized for the measurements.  Section III explains the selection
of reaction events and the methods used to determine
$\sigma_R$ and $\sigma_p$. The theoretical calculations are presented and
compared with our data in Section IV.   Section V briefly summarizes our
results and the conclusions we draw.  

\section{The Experiment }
Our measurements of $\sigma_R$ are similar to those made earlier for the He
and Li isotopes on Si \cite{rew96}.  Collimated monoenergetic projectiles
are selected in a multi-detector Si telescope. Reaction events are then
identified as those whose total energy loss in the telescope differs from
that of non-reacting projectiles. Since the projectile's energy decreases as
it travels through the telescope,  we obtained cross sections at different
energies by identifying the detector in which a reaction occurred; i.e., the
first detector to give a signal different from projectile signals.  

\begin{figure*}[tbh]
\begin{center}
\mbox{\epsfig{file=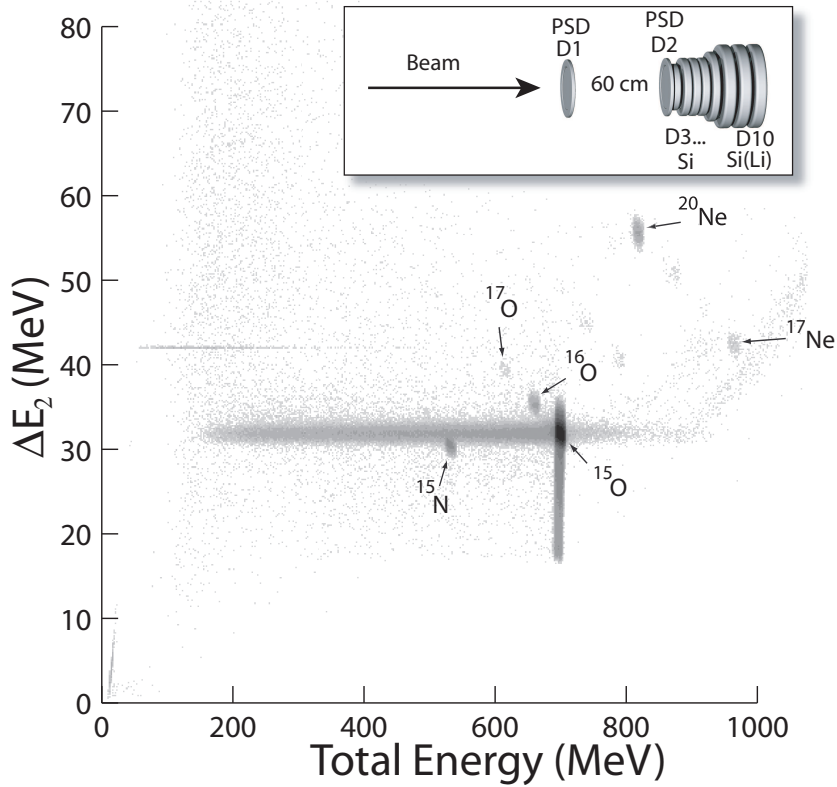,width=12cm}}
\end{center}
\caption{(color online) Scatter plot of $\Delta$E$_2$ vs. total energy deposited in the telescope, for
incident 48 MeV/nucleon $^{15}$O.  Most contaminants ($^{16,17}$O, $^{17}$Ne,
etc.) are removed by gating on  $\Delta$E$_2$; the treatment of $^{15}$N is
described in the text.  The inset shows the target-detector telescope used for
measurements. }
\label{fig1}
\end{figure*}

To measure $\sigma_p$, the heavy fragments from proton removal reactions
were observed in particle identification spectra from two detectors -- the
one which stops the fragment and the one which precedes it -- following
those in which the breakup reactions occur \cite{rew04}. 

An 80 MeV/nucleon $^{20}$Ne beam at the National Superconducting Cyclotron
Laboratory was fragmented on a 300 mg/cm$^2$ $^9$Be target which produced
secondary beams.  These secondary beams were transported by the
A1200 analyzing system \cite{A1200} to the Si detector telescope. In this
system the beams were partially purified by polyethylene wedges, and their
energy dispersions were limited to $\leq$1\% FWHM by analyzing slits. For
example, one A1200 setting delivered usable quantities of $^{11}$C,
$^{12}$N, and $^{13}$O at energies of 41, 46, and 51 MeV/nucleon,
respectively. 

The Si target-detector telescope is shown schematically as an inset to
Figure 1.  The first two detectors, which we call D1 and D2, were
position-sensitive detectors (PSD's), with thickness 0.2 mm, lateral dimensions 2.5 cm
x 2.5 cm, and 60 cm separation.  They measured projectile coordinates normal
to the beam axis, defining  beam of 5 mm radius.  Energy losses
in these detectors identified the desired projectiles, but allowed some contaminants as
we discuss later. The next five detectors, called D3 through D7, were about
0.5 mm thick and had active areas from 300 to 600 mm$^2$. They stopped all
non-reacting projectiles with A$\geq$11 and served as targets for reactions,
including breakup. The final Li-drifted detector, D8,  had 5 mm thickness
and about 4 cm diameter.  It stopped projectiles with A$\leq$10 and the
longer-ranged breakup fragments. All detectors except D1 were close-packed. 
Detectors D9 and D10 were not used in this experiment, but were utilized in
a concurrent measurement of $\sigma_R$'s in the Be isotopes \cite{rew01} .

The scatter plot of Fig.~\ref{fig1} illustrates our secondary beam composition.  It
shows $\Delta$$E_2$, the energy deposited in D2, vs. the total energy loss
in the telescope, for properly-aimed projectiles tentatively identified as
$^{15}$O by their energy loss in D1.    The non-reacting projectiles are the
most dense group, labeled $^{15}$O.  Events in the horizontal tail to the
left react downstream of D2, and the few to the right are either pileup or
positive Q-value reactions.  Events in the vertical tail are non-reacting
projectiles with significant energy straggling in D2. 

Several contaminant groups, with magnetic rigidities in the A1200 system identical to
that of 48 MeV/nucleon $^{15}$O, also are observed in Fig.~\ref{fig1}. Most of these,
e.g. $^{16,17}$O and $^{17,20}$Ne, are so well resolved  that a gate on
$\Delta$$E_2$ removes them completely.  The treatment of the remaining
$^{15}$N contaminant is described in the next section.

\section{Event Selection and Data Analysis}

\subsection{$\sigma_R$ Measurements}

Spectra of total energy deposited in the Si telescope by incident 48
MeV/nucleon $^{15}$O projectiles appear in Fig.~\ref{fig2}.  Each of these four
spectra shows a prominent peak for non-reacting projectiles and, below this
peak, a continuum due to reactions.

Events in the spectrum labeled D$_{1-2}$ are from projectiles with 
acceptable positions and normal $^{15}$O energy losses in Detectors D1 and
D2.  The $^{15}$O reactions beneath the $^{15}$N contaminant peak are
extracted by assuming a linear energy dependence of the $^{15}$O reaction
yield near that peak.  The contribution from small Q-value reactions is included by
extrapolating the $^{15}$O reaction yield just below the full energy peak to
its center.  The probability $\eta$$_2$ of a reaction occurring anywhere
downstream of Detector D2 then equals the ratio of reactions to total events
in this spectrum, which are mainly in the sharp peak.  The spectrum labeled D$_{1-3}$ has an additional
gate requiring normal $^{15}$O ionization in D3, and therefore gives the probability
$\eta$$_3$ for a reaction to occur beyond that detector.  Since a
non-reacting $^{15}$O projectile drops in energy from 44 to 38 MeV/nucleon
in Detector D3,  the average cross section for this energy range is 

\begin{equation}
\sigma_R = (\eta_2 -\eta_3)/N_3                           
\end{equation}
where Detector D3 has N$_3$ target nuclei per unit area.  Similarly, the
$\sigma_R$'s for the energy ranges 38 to 31 MeV/nucleon and 31 to 22
MeV/nucleon are determined by ($\eta_3 -\eta_4$)$/$N$_4$ and ($\eta_4 -\eta_5$)$/$N$_5$,
respectively.  Corrections were made for attenuation of the beam in the
individual detectors; these were less than 1\% in all cases.

Our $\sigma_R$ data are presented in Table~\ref{tabsigr}.  The main experimental
uncertainties are statistical in origin: the counts in the reaction region,
subtraction of contaminant events, and extrapolation to the
non-reacting-projectile peak center in Fig.~\ref{fig2}.    Events above the peak are
believed to be mainly pileup.  If treated similarly to those below the peak,
they would add about 1.5\%, or .02 b, to the $\sigma_R$'s quoted in Table~\ref{tabsigr}.
 Given the uncertain origin of these events, we make no correction to
$\sigma_R$ but add 2\% uncertainty in quadrature to the experimental
uncertainties. A further 2\% uncertainty in the detector thicknesses is also
included.

The contaminants produced very few reactions which were counted in the
reaction continuum. For example, there are about 1700 non-reacting  $^{15}$N
events in the D$_{1-2}$ spectrum of Fig.~\ref{fig2}, compared with 3.2x10$^6$
$^{15}$O's.  Hence the reactions by contaminants add about 0.05\% to the
measured  $\sigma_R$'s.  No correction was made for this effect since it is
so small and the contaminants' reaction cross sections are not well known.

\subsection{$\sigma_p$ Measurements}

The heavy fragments from proton removal generally have longer ranges than
the projectiles which produce them.  For example, all projectiles with A
$\geq$ 11 stopped in Detector D7, but most produced some fragments which
stopped in D8 along with others which stopped in D7. The exception was
$^{13}$O which had detectable yield only for $^{12}$N stopping in D7. 
Projectiles with A$\leq$10, and all of their fragments, stopped in D8.
Particle identification (PID) was used to select fragments of interest. The
PID parameter for fragments stopping in Detector 7 is obtained from the
energy losses $\Delta E_6$ and $\Delta E_7$,   through the equation
\be
PID = A[(\Delta E_6 + \Delta E_7)^p - (\Delta E_7)^p],                      
\ee
where $p=1.73$ and the arbitrary constant $A$ is chosen to place the peaks of
interest in convenient channels.  A similar equation, relating energy losses
$\Delta E_7$ and $\Delta E_8$, identifies those fragments stopping in D8. 
\begin{figure*}[tbh]
\begin{center}
\mbox{\epsfig{file=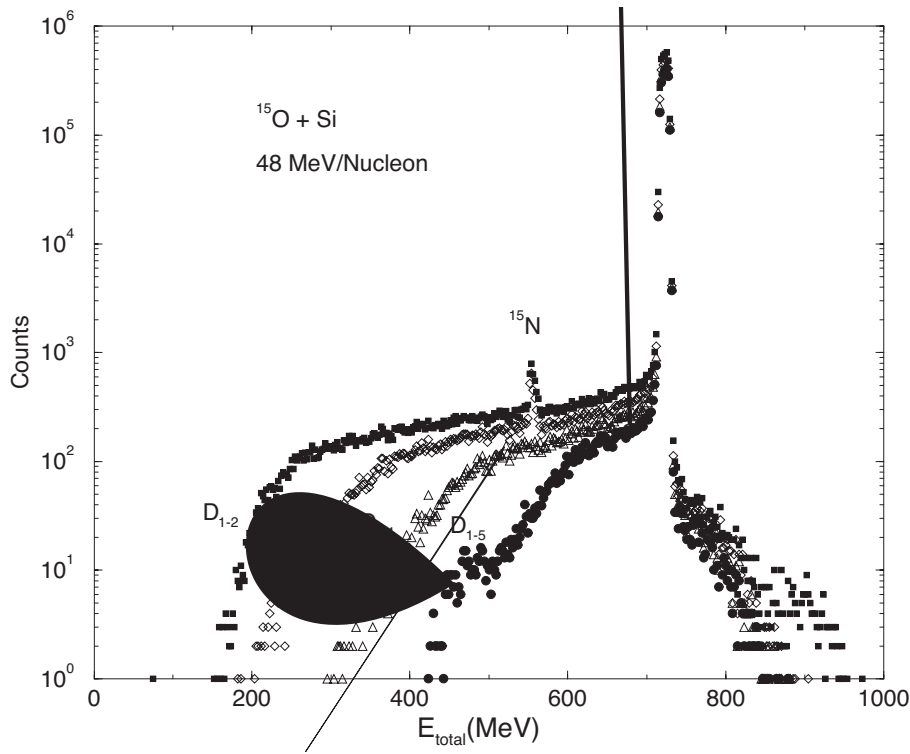,width=12cm}}
\end{center}
\caption{ 
Total energy spectra of 48 MeV/nucleon $^{15}$O incident upon the Si
telescope, gated to exclude events which react before detectors D3, D4, D5,
or D6.  The difference in reaction probabilities between curves labeled
D$_{1-2}$ and D$_{1-3}$ determines $\sigma_R$ at the energies the
incident projectiles have in D3,  after $^{15}$N contaminant events are
subtracted.}
\label{fig2}
\end{figure*}

Fragment identification was basically restricted to those produced in
Detectors D3 through D5.  A tight $\Delta$E$_2$ gate limited the production
region in D2 to about its last 0.03 mm, which increased the effective
production target thickness.  Fragments produced earlier in D2 were rejected
because of their decreased ionization.  Production in D6 and beyond was
excluded by counting only events with significantly lower $\Delta$E$_5$ than
that of non-reacting projectiles.  This rejected some events produced in D5
near its rear face, and we made a model correction for this loss.
\begin{figure*}[tbh]
\begin{center}
\mbox{\epsfig{file=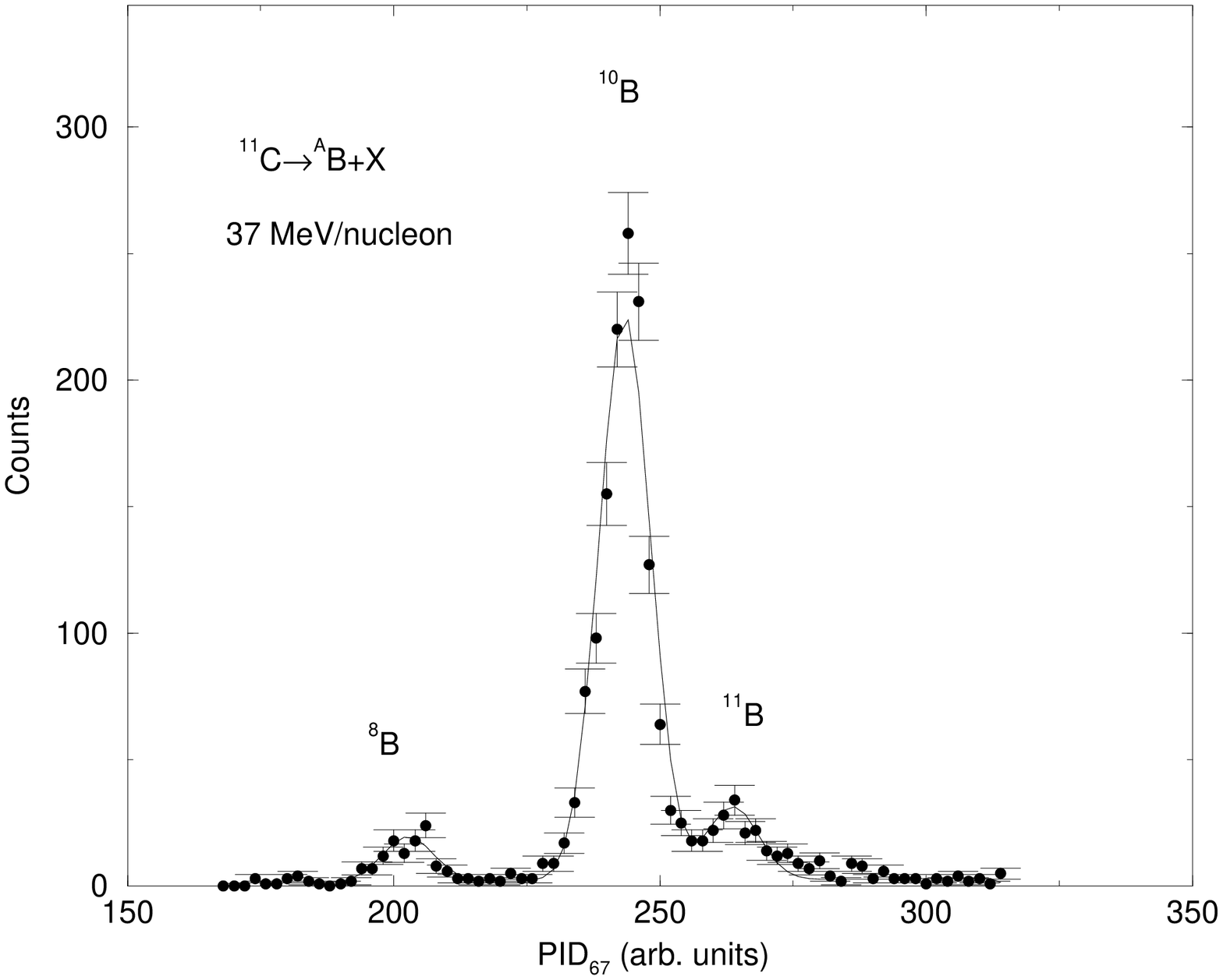,angle=-90,width=12cm}}
\end{center}
\caption{Particle identification spectra for bound B isotopes produced in
Detectors D3 through D5 and identified in D6 and D7, with 41 MeV/nucleon
$^{11}$C incident upon the telescope. Note the absence of a group due to $^9$B, which is unbound.}
\label{fig3}
\end{figure*}

As an example, we discuss the boron isotopes produced by incident 37
MeV/nucleon $^{11}$C. A PID spectrum for fragments stopping in D7 appears in
Fig.~\ref{fig3}.  The dominant isotope produced is $^{10}$B, with lesser amounts of
$^{11}$B (through charge exchange) and  $^8$B (via some more complex
process).  The gap between the $^8$B and $^{10}$B groups reflects the
absence of $^9$B, which is unbound.

Several corrections were made to the observed fragment yields. The largest
of these was for rejection of fragments produced in D5, near its rear face,
by the software gate on $\Delta$E$_5$ mentioned earlier.  This correction
required the fragment momentum distribution in the kinematic center of mass system (c.m) of the
incident projectile, which was calculated from Goldhaber's
\cite{rew04,goldhaber} theory. Typically, 10\% of the fragments stopping in
D7, but none of those stopping in D8, were rejected. No fragments produced
before D5 were rejected by this gate; thus, the overall event loss it caused
was typically 3\%. Since fragment production in D2 increased the effective
target thickness by about 2\%, the observed yield was reduced by this
amount.  The loss of projectiles by reactions in D3 through D5, added to
that of produced fragments lost to secondary reactions and hence
misidentified, was about 1\%. Possible corrections for fragments transmitted
through D8 or stopping short of D7 were found to be negligible.

The beam contaminants mentioned earlier had no effect on these measurements
due to their short range.  For example, the 37 MeV/nucleon $^{15}$N contaminant in the
48 MeV/nucleon $^{15}$O beam (see Figs.~\ref{fig1}~and~\ref{fig2}) stops in Detector D6.  In
contrast, the $^{14}$N fragments produced from $^{15}$O breakup reached at
least D7. 

The ratio of the combined corrected fragment yield to the incident flux
(projectiles with acceptable positions and energy losses in D1 and D2) gave
the breakup cross section; these are presented in Table~\ref{tabsigbu}.  The uncertainties
listed therein were found by adding in quadrature the statistical
uncertainties and an uncertainty of 1/3 of the combined corrections to the
data.

\section{Theoretical Predictions, and Comparison  with Measurements}
\subsection{Reaction cross sections}
\label{reac}
\begin{figure*}[tbh]
\begin{center}
\mbox{\epsfig{file=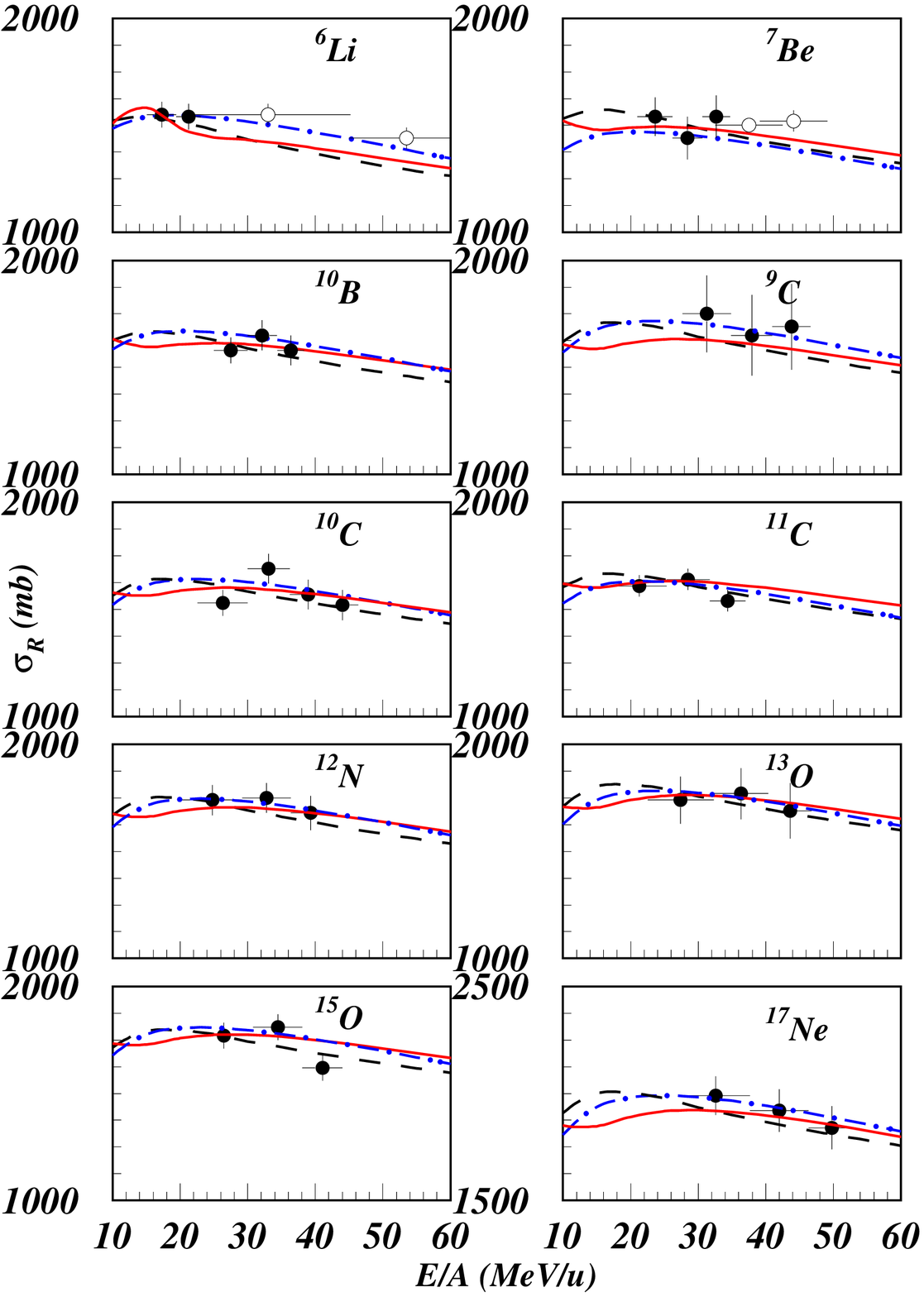,width=12cm}}
\end{center}
\caption{ (color online) Experimental reaction cross sections (filled points) compared
with predictions from the  Kox phenomenological formula (dash-dotted lines), the 
optical limit of the
Glauber model (dashed lines) and the JLM microscopic folding model (continuous
lines). Open symbols are data from our earlier work \protect\cite{rew96,rew01}. See 
text for details.  }
\label{fig4}
\end{figure*}

Experimental reaction cross section data are shown in Fig. \ref{fig4}.
We first compare these data with the predictions of the  strong absorption model of
Kox \cite{kox}. In this model the reaction cross section is
\be
\begin{array}{rcl}
\sigma_R(E) & = & \pi r_0^2\left(A_p^{1/3}+A_t^{1/3}+
a\frac{A_p^{1/3}A_t^{1/3}}{A_p^{1/3}+A_t^{1/3}}-C(E)\right)^2 \\
& &\times\left(1-\frac{V_c}{E_{c.m.}}\right)
\end{array}
\label{eqf1}
\ee
where $A_{p(t)}$ are the projectile (target) mass numbers, $a=1.85$ is a mass
symmetry parameter related to the volume overlap of projectile and target, and
$C(E)$ is a correction related to the transparency of the  optical potential.
We adopt here the linear approximation of Mittig \etal \cite{mittig},
$C(E)=0.31+0.0147E/A$ which gives reasonable results for the range of energies
studied here. Bending of 
trajectories in the target Coulomb field is taken into account by the last factor in Eq.~\ref{eqf1}. 
There, the Coulomb potential is evaluated at $R_c = \left[ 1.07 \left( A_p^{1/3} +A_t^{1/3}\right) +2.72 \right]$~fm.

The Kox formula gives excellent results
 for stable
nuclei when the reduced strong absorption radius is fixed at $r_0$=1.1 fm,
 and therefore any significant departure from its predictions may disclose a 
 halo structure.
The calculations shown in Fig. \ref{fig4} were done using the standard value for $r_0$,  except for $^6$Li and $^9$C where a somewhat larger value 
($r_0$=1.15 fm) was needed to fit the data. This variation must relate to the weak binding of the last two nucleons in each of these nuclei. It should be noted that the variation $\Delta
r_0$=0.05 fm adds 130 mb to $\sigma_R$ for $^6$Li at 30 MeV/nucleon.

\begin{figure*}[tbh]
\begin{center}
\mbox{\epsfig{file=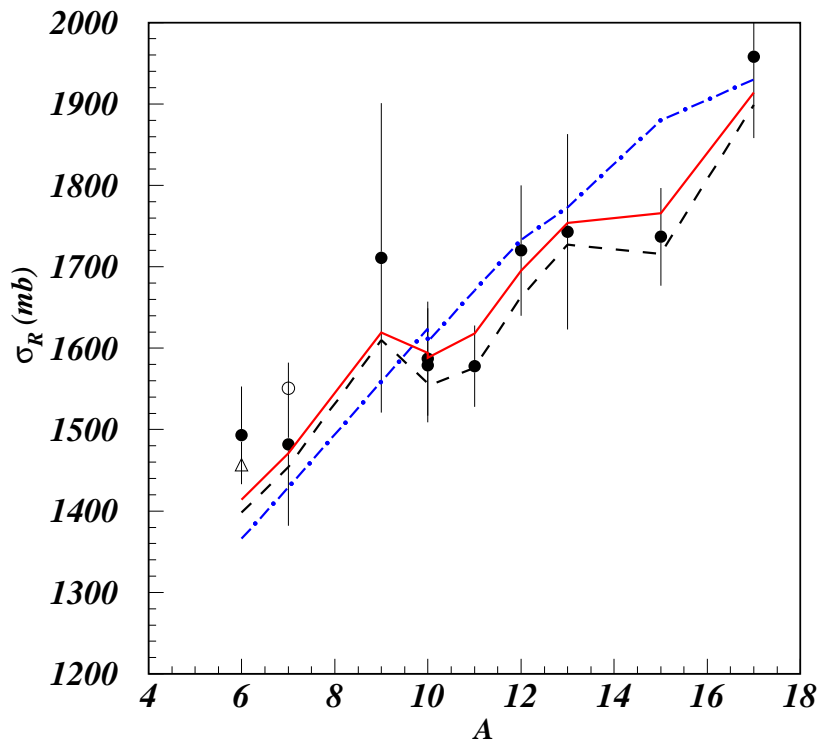,width=12cm}}
\end{center}
\caption{(color online) Interpolated reaction cross sections at 35 MeV/nucleon
(points) as a function of projectile mass number. Theoretical predictions use the
Kox model with the standard reduced strong absorption radius $r_0$=1.1 fm (dash-dotted
line), Glauber (OL) model (dashed line) and JLM model (full line). Open symbols are 
optical model
values from elastic scattering of $^6$Li+$^{28}$Si at 318 MeV (triangle,
\protect\cite{nadas318}) and from $^7$Li+$^{28}$Si at 350 MeV (circle,
\protect\cite{nadas350}).}
\label{fig5}
\end{figure*}

In Figure~\ref{fig5} we display the A-dependence of the Kox $\sigma_R$ predictions, using $r_0 = 1.10$~fm for all projectiles, at an energy of 35 MeV/nucleon. To compare them with our measurements, predictions for all projectiles were individually renormalized to best fit the energy dependence of each projectiles's $\sigma_R$ data. The renormalized calculations at 35 MeV/nucleon are plotted as interpolated ``data'' in Figure~\ref{fig5}.
The smooth $A$ dependence of the predictions reproduces the data on average, 
but is unable 
to explain the scatter of the reaction cross sections observed in the experiment. This observation suggests that a more sophisticated theory
 is needed to explain these data.

We next use the optical limit (OL) of the Glauber multiple scattering theory. 
The reaction
cross section is given by
\be
\label{eqf2}
\sigma_R=2\pi\int_0^{\infty}b [1-T(b)] db,
\ee
where $b$ is the impact parameter, and the transparency function (or elastic
survival probability) $T(b)$ is given
by
\be
\label{eqf3}
T(b)=\exp[-\chi (b)].
\ee
The quantity $[1-T(b)]$ is called the opacity or nuclear shadow.  The scattering phase 
is given by
\be
\label{eqf4}
\chi(b)=\sum_{\alpha,\beta=p,n}\sigma^{NN}_{\alpha\beta}(E)\int d\vec b_1 d \vec
b_2\tilde{\rho}_{\alpha}(b_1)\tilde{\rho}_{\beta}(b_2)\delta(\vec b_1+\vec b-\vec
b_2)
\ee 
where the sum runs over all NN isospin channels. Here we assume a zero-range 
nuclear force and charge symmetry for the free NN cross sections ($\sigma_{pp}=\sigma_{nn}$). 
These are taken from the parametrization of John
\etal \cite{john}. We ignore in-medium effects, Pauli blocking, and Fermi
motion, and assume a purely imaginary forward NN scattering amplitude.  In principle, one should correct
Eq. \ref{eqf2} for Coulomb dissociation effects. Estimation of this mechanism (see next
subsection) leads to the conclusion that its contributions are negligibly small. They
amount to 2 mb for $^7$Be and 20 mb for $^{12}$N Coulomb breakup at 35 MeV/nucleon,
well within the experimental uncertainties. The profile functions $ \tilde{\rho}$ needed in Eq.~\ref{eqf4} are calculated by
Abel transformation of the ordinary particle densities $\rho$ in coordinate space,
\be
\label{eqf5}
\tilde{\rho}(b)=\int_{-\infty}^{+\infty}\rho(\sqrt{b^2+z^2})dz 
\ee
Correction 
for bending of the trajectories in the target Coulomb field
is introduced as follows. Eq.~\ref{eqf2} is rewritten as
\be
\label{eqc1}
\sigma_R=2\pi\int_0^{\infty}b[1-T(b')]db
\ee
where $b'=(1/k)\left(\eta+\sqrt{\eta^2+k^2b^2}\right)$ is the impact parameter for a
grazing Coulomb trajectory, $\eta$ is the Sommerfeld parameter, and $k$ the
wave number. After some algebra one obtains
\be
\label{eqc2}
b^2=b'^2\left(1-\frac{V_c(b')}{E_{c.m.}}\right).
\ee

Using $V_c(b')\approx V_c(R_c)$ with $R_c$ defined above, one obtains
\be
\label{eqc3}
\sigma_R=2\pi \left(1-\frac{V_c(R_c)}{E_{c.m.}}\right)\int_0^{\infty}b'[1-T(b')]db'
\ee
which also justifies the correction made in Eq.~(\ref{eqf1}).

The single particle densities used in this analysis were obtained from a standard spherical 
HF+BCS calculation using the density functional of Beiner and Lombard \cite{bl}.
The surface strength of the functional has been slightly adjusted to 
reproduce the known
experimental binding energies. The $rms$ radii from HF calculations are listed in Table
\ref{tabrms}. They show a surprisingly good agreement with experimental data
from high energy reactions \cite{ozawa} and those obtained from the present $\sigma_R$ data, as
described below, especially for the loosely bound nuclei $^7$Be, $^9$C, $^{12}$N,
$^{13}$O and $^{17}$Ne. The results with the OL model are displayed in Figs.
\ref{fig4} and \ref{fig5}. Clearly, the mass dependence of the reaction cross
section is well reproduced since this approximation incorporates
realistic densities.

In the remainder of this section we discuss the ability of the JLM folding model \cite{jlm}
to describe reaction cross sections
as a further check of our densities. We adopt
their nuclear matter approach which incorporates a
complex, energy- and density-dependent parametrization of the effective
interaction obtained in the Brueckner Hartree-Fock approximation from the Reid
hard-core NN potential. Studies of elastic scattering of $p$-shell nuclei
 \cite{flo,trache00} indicate that the absorptive component of the JLM potential
is realistic  for loosely bound nuclei and needs no renormalization
( $N_w\approx $1, see below ), while the real part needs a significant renormalization.
Here  we extend these studies much closer to the proton drip line.

In the JLM model the complex form factor for the optical potential is given by
\begin{equation}
U(R)=\int d\vec{r}_{p}d\vec{r}_{t}\rho _{p}(r_{p})\rho _{t}(r_{t})v(\rho
,E,s)  \label{eqjlm1}
\end{equation}%
where $v$ is the (complex) NN interaction, $\rho _{p(t)}$ are the single
particle densities of the interacting partners, 
$\vec{s}=\vec{r}_{p}+\vec{R}-\vec{r}_{t}$
is the NN separation distance between interacting nucleons and $\rho $ is
the overlap density. The effective NN interaction contains an isovector
component, which gives a negligibly small contribution for $p$-shell nuclei.
However, it is included here for convenience in conjunction with
appropriate single particle isovector densities.
The coupling of the entrance channel to the breakup and particle transfer reactions has been described by a dynamic polarization potential (DPP) which is strong and has complicated dependence on radius, mass, and energy \cite{mac}. To simulate the radial dependence of a DPP, and to increase the flexibility of the
folding potential we introduce a smearing function $h(r)$ to obtain our final
folding potential,
\be
\tilde U(R)=\int d\vec R'U(R')h(\vert \vec R-\vec R'\vert)
\ee
 The smearing function
$h(r)$
is taken as a normalized Gaussian \cite{jlm,trache00},
\begin{equation}
h(r)=\frac{1}{t^{3}\pi ^{3/2}}\exp (-r^{2}/t^{2})  \label{eqjlm2}
\end{equation}%
which behaves as a $\delta $-function for $t\rightarrow 0$, while for finite
$t $ values it modifies the $rms$ radius of the folding form factor by 
$r_{h}^{2}=(3/2)t^{2}$, leaving the volume integral unchanged. It turns out 
that the smearing
procedure described above is essential in simulating the complicated radial
dependence of the dynamic polarization potential \cite{flo}.
To be consistent with the JLM model we take the overlap density in Eq.
\ref{eqjlm1} to be given by
\begin{equation}
\rho =\left[\rho _{p}(\vec{r}_{p}+\frac{1}{2}\vec{s})\rho
_{t}(\vec{r}_{t}-\frac{1%
}{2}\vec{s})\right]^{1/2} 
\label{eqjlm3}
\end{equation}%
 This approximation is physically reasonable since the
overlap density approaches zero when one of the interacting nucleons is far
from the core, and approaches the nuclear matter saturation value for complete
overlap. We recall that the JLM model was developed to describe
the optical potential for a nucleon traversing nuclear matter, and its density
 dependence is defined for
densities not exceeding the saturation value in nuclear matter.

The model contains four parameters: two normalization constants $N_v, N_w$ and
two range parameters $t_v, t_w$. They have been fixed here  
close to standard
values for $p$-shell nuclei \cite{trache00,flo}: $N_v$=0.4, $N_w$=0.85 and 
$t_v=t_w=1.2\;\mathrm{fm}$. We assume that 
the energy dependence is weak, and use this set of parameters at all energies and for
all projectiles. We are merely interested in a general assessement of the JLM model
rather than fitting a particular cross section.
The results of this approach are shown with continuous lines in
Figs. \ref{fig4} and \ref{fig5}. Fig. \ref{fig5} also includes the
optical model cross sections  obtained by fitting known elastic 
scattering distributions
\cite{nadas318,nadas350}. One observes an even better
description of the scatter effect in the reaction cross sections.

\begin{figure*}[tbh]
\begin{center}
\mbox{\epsfig{file=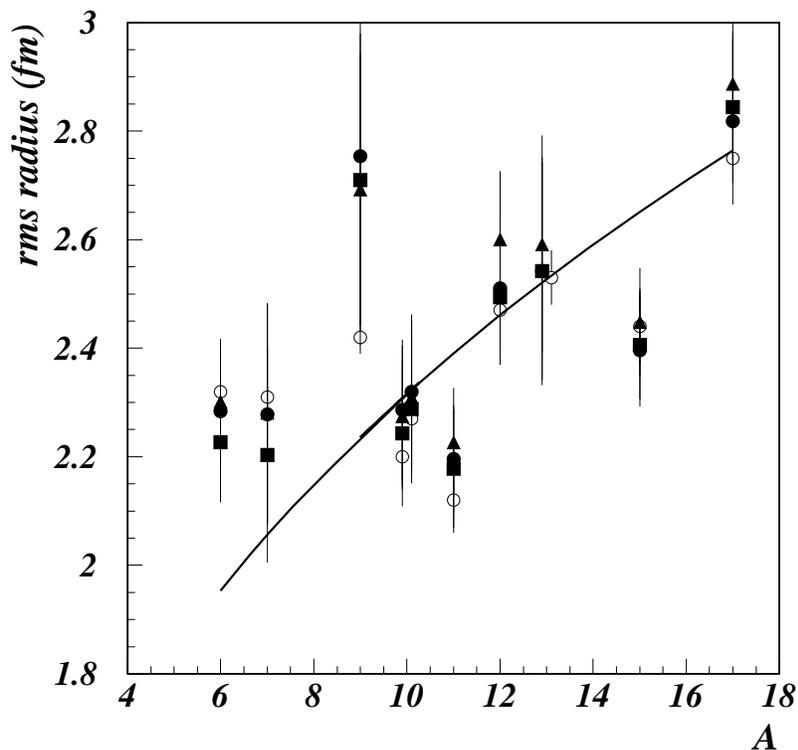,width=12cm}}
\end{center}
\caption{ $rms$ radii extracted from reaction cross sections measured in this experiment (filled symbols) using the gaussian density (filled circles), harmonic oscillator
density (squares) and Woods-Saxon density (triangles)) compared with the 
values extracted from high-energy  interaction
cross section \protect\cite{ozawa} (open circles). The line shown is 
1.075A$^{1/3}$fm to
guide the eye. Error bars are 1$\sigma$ deviation arising from energy
 dependence and statistics.}
\label{fig6}
\end{figure*}

\subsection{Nuclear radii}
It is interesting to use reaction cross section data to infer nuclear sizes. At
 the energies of our experiment, the total NN reaction cross section is large, the
mean free path is small, and
therefore the heavy ion reaction cross section is strongly influenced by NN collisions at the
surface. We chose to describe the tails of the nuclear matter densities by a
Gaussian\cite{karol}:
\be
\label{eqf6}
\rho_i(r)=\rho_{i0}\exp(-r^2/a_i^2),~i=p,t,
\ee
with the normalization,
\be
\label{eqf7}
\rho_{i0}=\frac{A_i}{(\sqrt{\pi}a_i)^3}.
\ee
The slope parameter $a_i$ is related to the nuclear size by
\be
\label{eqf8}
<r_i^2>=\frac{3}{2}a_i^2.
\ee
We let $a^2=a_p^2+a_t^2$ and
\be
\label{eqf9}
\chi_0=\overline{\sigma_{NN}}\pi^2\rho_{p0}\rho_{t0}\frac{a_p^3a_t^3}{a^2},
\ee
where $\overline{\sigma_{NN}}$ is the isospin-averaged NN cross section. Under these assumptions,
Eq. (\ref{eqf2}) can be solved analytically,
\be
\label{eqf10}
\sigma_R=\pi a^2\left(C+\ln\chi_0+E_1(\chi_0)\right)\left(1-\frac{V_b}{E_{c.m.}}\right)
\ee
where $C=0.5772$ is the Euler constant and $E_1$ is the exponential integral.
Eq. (\ref{eqf10}) shows essentially the geometric character   of the reaction 
cross
section since the leading term is $\sigma_R\approx\pi(r_p^2+r_t^2)$. The leading term in the Kox formula is $\sigma_R\approx\pi(r_p+r_t)^2$. The energy
dependence is governed entirely by $\sigma_{NN}$.  To extract nuclear sizes from experimental data, we use Eq.(\ref{eqf10}), with the
target $rms$ radius fixed to $r_t$=3.05 fm which we obtained from HF calculations.
 The compilation of Angeli \cite{angeli} for $rms$ charge radii
indicates a value of $r_{ch}$=3.12 fm for $^{28}$Si. We calculate an upper limit of 2$\%$ uncertainty from
this difference in our theoretical estimations. In practice, the 
parameter $a_p$ is gridded in small steps until
the calculated cross section equals the experimental value. The results are
listed in Table \ref{tabrms} and displayed in Fig. \ref{fig6}. These values are 
consistent with those extracted
from high energy data within their uncertainties. The values and uncertainties quoted in the 
table include
the weighted averages over various energies measured here, and statistical uncertainties. No provision has been taken to correct values in Table III for nucleon finite size. Assuming a nucleon rms value of $r^2_n\approx$0.8 fm$^2$, this would increase values in Table III by $\approx$5\%.

The sensitivity of the cross section to the functional form of the single-particle
density is studied using harmonic
oscillator wave functions appropriate for the $pd$-shell nuclei with $s$ and $d$ state admixtures.
\be
\label{eqnew1}
\rho_{\tau}(r)=\frac{1}{(\sqrt{\pi}b_{\tau})^3}\left(n_s+\frac{2}{3}n_p x^2+\frac{4}{15}n_d x^4\right)
\exp{(-x^2)}
\ee
where $n_s$, $n_p$ and $n_d$  are occupation numbers in the $s$, $p$ and $d$ shells,
$b_{\tau}$ is 
the range parameter and $x=r/b_{\tau}$. The finite range of the nuclear force is included in Eq.
  (\ref{eqf4}) by replacing the $\delta$ force by a finite range NN interaction
\be
\label{eqnew2}
v(\vec b)=\frac{1}{\pi \mu^2}\exp{(-b^2/\mu^2)}
\ee
with the interaction range  $\mu$=1 fm \cite{berci}. The normalization in Eq.
(\ref{eqnew2}) ensures that the reaction volume overlap in Eq. {\ref{eqf4} is
not changed, but the smearing enhances the weight of the density tail and reduces
the contribution from the central part of the density. In addition, the number
of NN inelastic scatterings in the overlap volume is weighted differently by
$\sigma_{pp}\ne\sigma_{np}$. As a result, the $rms$ radii extracted with this
method (shown in Table \ref{tabrms}) are slightly smaller than with the Karol
model. 
Woods-Saxon (WS) shapes   are more
appropriate for testing the role played by long tails in the projectile neutron and/or 
proton densities
\be
\label{eqnew3}
\rho_{\tau}(r)={\rho_{0\tau}}/\left[1+\exp{\left(\frac{r-R_{\tau}}{a_{\tau}}\right)}\right],~\tau=p,n
\ee
where the $\rho_{0\tau}$ are normalization constants. Equations
(\ref{eqf2}-\ref{eqf5}) have been solved numerically. The half-radius parameter
$R_{\tau}$
is the most sensitive in extracting the nuclear $rms$ radius. No anomalies were
found in proton or neutron surface thickness when using  $a_p\approx a_n\approx$0.5 fm
for most nuclei, and slightly larger values for $^9$C, $^{12}$N and $^{17}$Ne. The uncertainties shown in Table
\ref{tabrms} are evaluated with the bootstrap method of Efron \cite{efron},
which has the merit of weakening the influence of systematic errors in the data,
and leads 
to meaningful ranges in the extracted parameters. All these calculations lead to
mutually consistent $rms$ radii for the nuclei studied here (Table
\ref{tabrms}).

\subsection{One-proton removal}
\label{sec1p}
We use the core spectator model and Glauber multiple scattering theory to
calculate one-proton breakup cross sections. This model has been tested
extensively on a number of one-neutron removal reactions of neutron-rich nuclei
in the $psd$ shell \cite{sauvan}. We approximate the ground state of the
projectile ($J^{\pi}$) by a superposition of configurations
of the form $[I_c^{\pi}\otimes nlj]^{J^{\pi}}$, where $I_c$ denotes the core
states and $nlj$ are the quantum numbers specifying the single particle wave
function of the outermost proton. The s.p. wave functions were obtained for
a Woods-Saxon potential using the effective separation energy recipe
$S_p^{eff}=S_p+E_{ex}^c$, where $E_{ex}^c$ is the excitation energy of the 
core state.  The depth of the potential was varied in such a way that the known one-proton separation energy was reproduced. The radius of the WS potential was fixed 
to values close to
the $rms$ radius of the core (Table~\ref{tabrms}) and the diffuseness was fixed to $a=0.5$ fm for all
cases. 

Fine tuning of these parameters could significantly improve the 
theoretical 
results since they determine the asymptotic normalization coefficient of the 
outermost proton.
We
neglect dynamical excitation of core states in the reaction. In this
approximation, the reaction can populate core states only to the extent that
there is a nonzero spectroscopic factor $C^2S(I_c,nlj)$ in the projectile ground
state. The various configurations are assumed to contribute incoherently to the total
breakup cross section for a given core state,
\be
\label{eqbu1}
\sigma_{p}(I_c^{\pi}) =\sum_{nlj}C^2S(I_c^{\pi},nlj)\sigma_{sp}(nlj,S_p^{eff}).
\ee
The total breakup cross section ($\sigma_p$) 
 is then the sum over all particle-bound 
states of the core.
In the present experiment, the core states were not identified and the
removed  proton was not detected. Therefore, the dynamical factor $\sigma_{sp}$ in
Eq.(\ref{eqbu1}) includes contributions from stripping, nuclear dissociation and Coulomb
dissociation. Other more violent channels, such as core breakup, have been
ignored. The nuclear mechanisms have been described using transition operators
defined in terms of scattering functions (S-matrix) for p-target and core-target
interactions generated in the JLM folding model as described above. Scattering
functions in impact parameter representation have been calculated in the eikonal
approximation, including noneikonal corrections up to second order. Coulomb
dissociation has been calculated in the first order of  perturbation theory. Both
$E1$ and $E2$ amplitudes have been included. The interference of these
amplitudes does not contribute to the inclusive removal cross section.

The spectroscopic factors $C^2S(I_c,nlj)$ employed here were calculated with the
shell model code OXBASH \cite{oxb} using the WBP \cite{wbp} interaction within a
$1s-1p-2s1d-1f2p$ model space. Excitations up  to $4\hbar\omega$ have been included in this 
single particle model space for most of the
cases. Inclusion of higher $n \hbar \omega$ excitations with this 
effective interaction was 
shown to better describe some experimental observables \cite{u4hbw}.
Spurious center-of-mass components of the projectile have been suppressed by the usual method \cite{rcom}
of adding a center-of-mass hamiltonian to the nuclear interaction.
In all cases, the experimentally established 
spin-parity and core excitation energies have been used.
The shell model spectroscopic factors have been multiplied by the
center-of-mass correction factor $A_p/(A_p-1)$, following Refs. ~\cite{browncm,diep74}.

The projectile ground state spin-parity ($J^{\pi}$), core spin-parity and 
excitation energies
(E$_{ex}$, I$_c^{\pi}$), the stripping, diffractive dissociation and Coulomb 
dissociation cross section,
as well as the shell model spectroscopic factors are listed in 
Table~\ref{tabre}. The total inclusive
cross section $\sigma_{p}^{Glauber}$ was corrected for center-of-mass effects as 
explained above. All calculations were done at 35 MeV/nucleon.

Comparison with experimental values in Table \ref{tabsigbu} shows reasonable
agreement. It should be noted that the extended Glauber model is designed to
describe one nucleon removal reactions for loosely bound nuclei. The wave
function of the outermost nucleon is assumed to penetrate substantially
into the classically forbidden region, since most of the reaction probabilty is
localized at the surface. The interaction should be strongly absorptive. In addition, the core is assumed to be compact and well 
decoupled from the outermost nucleon. A special case is $^{10}$B which has a quite large one-proton separation energy  (S$_p\approx$ 6.8 MeV), and a relatively
fragile core, $^{9}$Be. In this case the reaction is not completely peripheral, with
significant contributions arising from impact parameters $b\le R_c$, where $R_c$ is the core
radius. The calculated breakup cross
section exceeds the experimental value by more than 50$\%$. Only by excluding contributions from the
nuclear interior can one obtain a reasonable value (Table \ref{tabre}).

\subsection{Discussion of Results}

The reaction cross sections for $^6$Li measured in this experiment and at higher energy \cite{rew96} show in Fig.~\ref{fig4} a smooth energy dependence, falling at 50 MeV/nucleon, which is well fitted by the Kox model and slightly underpredicted by the Glauber models. For $^7$Be we see a flat energy dependence, with all models slightly underpredicting the 40 MeV/nucleon datum.

Figures \ref{fig4} and  \ref{fig5} show a large reaction cross section for the
$^9$C projectile. Both Glauber (OL) and JLM models predict this increase,  in
reasonable agreement with the experiment. The $rms$ radius obtained from
our present data exceeds by about 12\%
the value obtained from high energy data \cite{ozawa}, and it is close to that obtained for $^{17}$Ne,
another loosely bound nucleus. The smaller Coulomb barrier in $^9$C enhances the
penetrability of the outermost proton much beyond the range of nuclear forces
and this compensates the larger interaction volume provided by $^{17}$Ne. 
Unfortunately, the uncertainty is relatively large and the halo structure of this nucleus
cannot be firmly established. More accurate experimental data are needed to clarify 
this point.

The reaction cross sections for $^{10}$B, $^{10}$C, and $^{11}$C are definitely smaller than for $^9$C and not much larger than those of $^6$Li and $^7$Be. The extracted $rms$ radius of $^{11}$C is quite small though in agreement with high energy data \cite{ozawa}.  An independent measurement by Liatard \etal \cite{liatard} indicates a much
larger value, 2.46$\pm$0.30 fm, in agreement with our HF calculation of 2.48 fm (see Table \ref{tabrms}). 
$^{12}$N seems to behave normally. Both Glauber (OL) and JLM models give an
excellent description of the cross section. The extracted radii seem to be in good agreement
with HF calculations (see Table \ref{tabrms}). 

Figure~\ref{fig5} shows that $\sigma_R$ stays nearly constant from $^{12}$N through $^{15}$O, as predicted by the JLM and OL (but not the Kox) models. For well-bound projectiles we expect an increase with increasing A. However, $S_p$ is 0.6, 1.5 and 7.3 MeV for $^{12}$N, $^{13}$O and $^{15}$O, respectively, and the weaker binding of the two lighter nuclei increases the range of their valence protons. The two effects appear to roughly compensate for each other.

 Among the nuclei we studied, $^{12}$N has by far the largest one proton removal reaction cross section. Furthermore, the $p$-removal process for this case is related to $^{11}$C+$p$
radiative capture.  Depending on their initial CNO abundances, this reaction may have been important in some super-massive stars in the early universe, allowing the stars to explode as supernovae rather than collapsing as black holes before ejecting any mass.\cite{fuller86,wiescher89}.  The
asymptotic normalization coefficient (ANC) for $^{12}$N $\rightarrow$
$^{11}$C+p has been measured recently using transfer reactions \cite{tang}. 
However, these reaction-rate data are ambiguous and therefore it is
desirable to remeasure the ANC by breakup, as an independent check.
The one proton separation energy is small ($S_p=0.6
$ MeV), and thus $^{12}$N could be a proton halo candidate. The Glauber model nicely 
reproduces the
measured $\sigma_p$. However, the shell model calculations suggest a very 
fragmented structure
for the g.s. wave function, significant contributions arising from 
(g.s, 3/2$^-$$\otimes$$1p_{1/2}$), (g.s, 3/2$^-$$\otimes$$1p_{3/2}$)
(E$_x$=2.0 MeV, 1/2$^-$$\otimes$$1p_{3/2}$)
and (E$_x$=4.8 MeV, 3/2$^-$$\otimes$$1p_{1/2}$). The ANC's have been measured \cite{tang}, for the core ground state
components; they are $C^2_{1p_{1/2}}=1.4\pm0.2$ fm$^{-1}$ and
$C^2_{1p_{3/2}}=0.33\pm0.05$ fm$^{-1}$. Since the single 
particle normalization coefficients are almost identical for the $1p_{1/2}$ and 
$1p_{3/2}$ wave functions,
one should have approximatively 
$S_{1p_{1/2}}/S_{1p_{3/2}}\approx C^2_{1p_{1/2}}/C^2_{1p_{3/2}}$, a relation that is satisfied for the states mentioned above for which the experimental values of the ANC are known, when one uses the corresponding shell model spectroscopic factors listed in Table~\ref{tabre}.
The reaction cross section measurements showed no anomaly
for this nucleus. The calculated parallel and transverse core fragment momentum distributions are 
narrow (FWHM$_z\sim 86$ MeV/c, FWHM$_x\sim 120$ MeV/c) comparable to that found
 for a well
established halo nucleus, such as $^8$B \cite{kelley}. The halo character of $^{12}$N
could not be firmly established on the basis of the present data. More precise
reaction and breakup cross section data are needed as well as a clear separation of
contributions from core excited states.

For halo nuclei, one can assume that the core is decoupled from the halo nucleon, and the
following decomposition of the absorption operator (Eq. \ref{eqf2}) holds
\be
\label{eqd1}
1-T\approx 1-T_cT_h=(1-T_c)+T_c(1-T_h)
\ee
Here, the first bracketed term on the right describes the core absorption 
(in the absence of the halo particle), while the second describes the absorption 
of the halo particle, weighted by the core survival probability. One can associate 
this last term with the stripping component of the total breakup cross section $\sigma_{p}$. 
Assuming that the diffractive dissociation and Coulomb dissociation  components are 
small, one can  approximate $\sigma_R=\sigma_R^c+\sigma_{p}$. 
The difference of the $\sigma_{\mathrm{R}}$'s given for $^{11}$C and $^{12}$N in the middle 
energy bins of Table \ref{tabsigr} (at energies 28 and 33 MeV/nucleon, respectively) is 110 $\pm$ 
86 mb. This agrees with the breakup cross section of 120 $\pm$ 6 mb given in 
Table \ref{tabsigbu}, though with a large uncertainty.

$^{15}$O has a one-proton separation energy $S_p$ four times larger 
than $^{13}$O, yet
the $^{15}$O breakup cross section is two times larger. We expect a larger $S_p$
 to give a smaller $\sigma_p$  and so the Glauber model, which is very
sensitive to the separation energy, is not able to reproduce this behaviour.
This anomaly probably results from competition with $2p$-breakup; we
found that the $^{13}$O breakup produces more C than $^{12}$N, though the C yield is
isotopically unresolved. Other similar cases were observed for nuclei which have
one nucleon outside a fragile core. For eaxmple, $^{12}$Be, which has one neutron
loosely bound to a neutron halo nucleus $^{11}$Be, has a smaller $\sigma_n$ than 
$\sigma_{2n}$ \cite{rew01}. $^{9}$C behaves similarly for proton removal
\cite{rew04}.

\section{Conclusions}
We have measured reaction cross sections  $\sigma_R$ and one-proton removal cross sections
$\sigma_p$ for a range of stable and short-lived nuclei close to the proton drip line.  These $\sigma_R$  extend our earlier
measurements to projectiles of higher A and Z. Comparison with earlier, higher energy data for
$^6$Li and $^7$Be shows a reasonable energy dependence. The optical limit of the
multiple scattering
Glauber model, and double folding optical potentials derived from the NN effective interaction in nuclear matter, are about equally successful in describing the data,
including their energy- and mass-dependence. However, more accurate data would be
needed to distinguish between them or to test sensitively for nuclei with weak halos. 

Root-mean-square  nuclear radii have been
extracted using Glauber theory and three different functional forms for the
projectile
single particle densitities. These results are mutually consistent and in good agreement
with values extracted from high energy data, except for $^9$C in which case a
significantly larger radius has been found. 

One proton removal cross sections vary
widely. These reactions were
described within an extended Glauber model by incorporating fundamental NN
interactions and spectroscopic factors from the shell model.  The large $\sigma_p$ in the 
$^{12}$N case suggests a possible weak halo, although the wave function is fragmented into
many components. An anomaly has been
found in the A=12-15 region, where the one-proton removal cross sections and
total reaction cross sections do not behave as expected from proton
separation-energy systematics. More experimental and theoretical effort shoud be
devoted to a better understanding of the competition between $-1p$ and $-2p$
channels for nuclei, such as $^{13}$O, where the last proton's separation energy exceeds that of the next one (i.e.,$S_p > \frac{1}{2}S_{2p}$).

\section{acknowledgments}
We benefitted from useful discussions with Shalom Shlomo.
We thank J. J. Kruse, M. Y. Lee, T. W. O'Donnell, P. Schwandt, H. Thirumurthy,
J. Wang, J. Woodroffe, and J. A. Zimmerman for assistance with these
measurements.  One of us (F.C.) acknowledges support by the Texas A\&M
University Cyclotron Institute. Additional support from the following National Science Foundation
grants PHY-9971836 (UM-Dearborn), PHY02-44989 (Notre Dame and UM-Ann
Arbor, PHY0244453 (Central Michigan University) is acknowledged. B. Davids also acknowledges additional support from the Natural Sciences and Engineering Research Council of Canada.

\pagebreak

\pagebreak

\section{tables}
\label{tabsec}
\begin{table}[htbp]
\caption{Total reaction cross sections, $\sigma_R$,  for ten light isotopes on Si, averaged between the listed energies.}

\begin{center}
\begin{tabular}{||c|c|c||}
\hline
System       & Energy interval     & $\sigma_R$ \\  
             & (MeV/nucleon) & (b)      \\ \hline
$^6$Li+Si    & 15.1-19.4  &1.55$\pm$0.06\\
             & 19.4-23.1  &1.54$\pm$0.06\\ \hline
$^7$Be+Si    & 21.0-26.2  &1.54$\pm$0.09\\
             & 26.2-30.6  &1.44$\pm$0.10\\ 
             & 30.6-34.7  &1.54$\pm$0.10\\ \hline
$^{10}$B+Si  & 25.0-30.0  &1.58$\pm$0.06\\
             & 30.0-34.3  &1.65$\pm$0.07\\ 
             & 34.3-38.4  &1.58$\pm$0.07\\ \hline
$^9$C+Si     & 27.7-34.9  &1.75$\pm$0.18\\
             & 34.9-41.0  &1.65$\pm$0.19\\ 
             & 41.0-46.6  &1.69$\pm$0.20\\ \hline
$^{10}$C+Si  & 22.6-30.0  &1.53$\pm$0.06\\
             & 30.0-36.2  &1.69$\pm$0.07\\ 
             & 36.2-41.7  &1.57$\pm$0.07\\ 
             & 41.7-46.4  &1.52$\pm$0.07\\ \hline
$^{11}$C+Si  & 17.3-25.3  &1.61$\pm$0.05\\
             & 25.3-31.7  &1.64$\pm$0.05\\ 
             & 31.7-37.0  &1.54$\pm$0.05\\ \hline
$^{12}$N+Si  & 20.3-29.2  &1.74$\pm$0.07\\
             & 29.2-36.4  &1.75$\pm$0.07\\ 
             & 36.4-42.2  &1.68$\pm$0.08\\ \hline
$^{13}$O+Si  & 22.5-32.3  &1.74$\pm$0.11\\
             & 32.3-40.4  &1.77$\pm$0.12\\ 
             & 40.4-46.8  &1.69$\pm$0.13\\ \hline
$^{15}$O+Si  & 22.1-30.8  &1.77$\pm$0.06\\
             & 30.8-38.1  &1.81$\pm$0.06\\ 
             & 38.1-44.0  &1.62$\pm$0.06\\ \hline
$^{17}$Ne+Si & 27.5-37.7  &1.99$\pm$0.09\\
             & 37.7-46.3  &1.92$\pm$0.10\\ 
             & 46.3-53.3  &1.84$\pm$0.10\\ \hline
\end{tabular}  \\
\end{center}
\label{tabsigr}
\end{table}

\begin{table}[htbp]
\caption{Proton-removal cross sections for light nuclei on Si, averaged
between the specified energies. 
The Glauber theoretical predictions are described in Section~\ref{sec1p}.}

\begin{center}
\begin{tabular}{||c|c|c|c|c||}
\hline
Projectile,    &   Fragment   &    Energy interval   &  Measurement & Glauber\\
cross section  &              & (MeV/nucleon) & (mb) & (mb)\\
 \hline
$^7$Be,   $\sigma_{1p}$  &  $^6$Li    &  26-38 &  90$\pm$6  & 117\\
$^{10}$B, $\sigma_{1p}$  &  $^9$Be    &  30-42 &  41$\pm$3  & 62\\
$^{11}$C, $\sigma_{1p}$  &  $^{10}$B  &  17-37 &  53$\pm$2  & 78\\
$^{12}$N, $\sigma_{1p}$  &  $^{11}$C  &  20-42 & 120$\pm$6  & 130\\
$^{13}$O, $\sigma_{1p}$  &  $^{12}$N  &  23-49 &  31$\pm$5 & 56 \\
$^{15}$O, $\sigma_{1p}$  &  $^{14}$N  &  22-44 &  64$\pm$3  & 46\\
$^{17}$Ne,$\sigma_{2p}$  &  $^{15}$O  &  28-53 & 223$\pm$18 &\\ 
\hline
\end{tabular}
\end{center}
\label{tabsigbu}
\end{table}

\begin{table}[htbp]
\caption{Proton (r$_{\pi}$), neutron (r$_{\nu}$), 
and matter (r$_{\mathrm{m}}$) rms radii from HF calculations are given in Columns  2 through 4.  
Column 5 gives experimental values extracted from high-energy reaction cross sections \cite{ozawa}. 
The last three columns show experimental values extracted from our present data, using 
Glauber theory (OL) and assuming gaussian, harmonic oscillator, and Woods-Saxon densities.}

\begin{center}
\begin{tabular}{||c|c|c|c|c|c|c|c||}
\hline

Nucleus    &   $r_{\pi}$   &    $r_{\nu}$   &  $r_m$
&$r_m$(exp) \protect\cite{ozawa}&$r_m$(gauss)&$r_m$(HO)&$r_m$(WS)\\
\hline
$^6$Li    &  2.33 &    2.31   &  2.32& 2.32$\pm$ 0.03&2.28$\pm$ 0.10&2.23$\pm$0.11&2.30$\pm$0.12\\
$^7$Be    &  2.49 &    2.24   &  2.39& 2.31$\pm$ 0.02&2.28$\pm$ 0.20&2.20$\pm$0.19&2.28$\pm$0.20\\
$^{10}$B  &  2.46 &    2.46   &  2.46& 2.20$\pm$ 0.06&2.29$\pm$ 0.13&2.24$\pm$0.13&2.27$\pm$0.13\\
$^{9}$C   &  2.76 &    2.25   &  2.60& 2.42$\pm$ 0.03&2.75$\pm$ 0.34&2.71$\pm$0.26&2.69$\pm$0.25\\
$^{10}$C  &  2.57 &    2.31   &  2.47& 2.27$\pm$ 0.03&2.32$\pm$ 0.14&2.29$\pm$0.14&2.31$\pm$0.14\\
$^{11}$C  &  2.53 &    2.41   &  2.48& 2.12$\pm$ 0.06&2.20$\pm$ 0.10&2.18$\pm$0.11&2.23$\pm$0.10\\
$^{12}$N  &  2.70 &    2.45   &  2.60& 2.47$\pm$ 0.07&2.51$\pm$ 0.13&2.49$\pm$0.12&2.60$\pm$0.12\\
$^{13}$O  &  2.81 &    2.47   &  2.69& 2.53$\pm$ 0.05&2.54$\pm$ 0.21&2.54$\pm$0.20&2.59$\pm$0.20\\
$^{15}$O  &  2.69 &    2.58   &  2.64& 2.44$\pm$ 0.04&2.40$\pm$ 0.10&2.40$\pm$0.10&2.45$\pm$0.10\\
$^{17}$Ne &  2.98 &    2.61   &  2.83& 2.75$\pm$ 0.07&2.82$\pm$ 0.15&2.84$\pm$0.14&2.89$\pm$0.14\\
\hline

\end{tabular}
\end{center}
\label{tabrms}
\end{table}

\begin{table}[htbp]
\begin{center}
\caption{Calculated spectroscopic factors (C$^2$S) and cross sections 
$\sigma$(I$_c^{\pi}$) to the core excited states (E$_{ex}$,I$_c^{\pi}$)
populated in single proton removal from the projectile ($^A$Z,J$^{\pi}$) by the
silicon target. The contribution arising from stripping $\sigma_{abs}$,
diffractive dissociation ($\sigma_{diff}$) and Coulomb
dissociation($\sigma_{coul}$) are detailed. The total inclusive cross section
$\sigma_{p}^{Glauber}$ is corrected for center-of-mass effects as explained in the main
text.}
\noindent \begin{tabular}{|c c c c c c c c c c|}
\hline
\hline
 $^A$Z & $J^{\pi}$ & E$_{ex}^c$  & I$^{\pi}_c$ & $nlj$  & $C^2S$ & $\sigma_{abs}$ & $\sigma_{diff}$ & $\sigma_{coul}$&$\sigma(I^{\pi}_c)$\\
  & & [MeV]  &  &   &  & [mb] & [mb] & [mb]& [mb]\\
\hline
\hline
 $^{7}$Be & $3/2^-$ & g.s.  & $1^+$ & 1p$_{3/2}$ & 0.368 & 20.3 & 16.9 & 1.0 &38.2 \\
          &         &         &     & 1p$_{1/2}$ & 0.306 & 16.0 & 13.1 & 0.8 &29.9 \\ 
          &         & 2.186 & $3^+$ & 1p$_{3/2}$ & 0.365 & 18.0 & 14.3 & 0.5 &32.8  \\
\hline
\multicolumn{10}{|r|}{$\sigma_{p}^{Glauber}$=117 mb}\\
\hline
 $^{10}$B & $3^+$ & g.s.  & $3/2^-$ & 1p$_{3/2}$ & 1.097 & 29.3 & 24.6 & 2.30&56.2 \\
\hline
\multicolumn{10}{|r|}{$\sigma_{p}^{Glauber}$=62 mb}\\
\hline
 $^{11}$C & $3/2^-$ & g.s.  & $3^+$ & 1p$_{3/2}$ & 0.927 & 16.8 & 13.5 & 0.8&31.1 \\
          &         & 0.718 & $1^+$ & 1p$_{3/2}$ & 0.756 & 13.2 & 10.5 & 0.5&24.2  \\
          &         &         &     & 1p$_{1/2}$ & 0.532 &  8.6 &  6.8 & 0.3&15.8 \\ 
\hline
\multicolumn{10}{|r|}{$\sigma_{p}^{Glauber}$=78 mb}\\
\hline
 $^{12}$N & $1^+$ & g.s.  & $3/2^-$ & 1p$_{3/2}$ & 0.073 & 3.3 & 3.4 & 2.0  &8.8 \\
          &         &         &     & 1p$_{1/2}$ & 0.518 & 21.6& 22.5& 13.5 &57.7 \\ 
          &         &         &     & 2p$_{3/2}$ & 0.001 & 0.07& 0.07& 0.04 &0.2 \\ 
          &         &         &     & 2p$_{1/2}$ & 0.002 & 0.13& 0.14& 0.08 &0.35 \\ 
          &         & 2.000 &$1/2^-$& 1p$_{3/2}$ & 0.302 & 8.9 & 8.5 & 2.3  &19.7  \\
          &         &         &     & 1p$_{1/2}$ & 0.038 & 1.0 & 0.9 & 0.3  &2.2 \\
	  &         & 4.318 &$5/2^-$& 1p$_{3/2}$ & 0.130 & 2.9 & 2.6 & 0.40 &5.9  \\ 
          &         & 4.800 &$3/2^-$& 1p$_{3/2}$ & 0.085 & 1.8 & 1.6 & 0.2  &3.6  \\
          &         &         &     & 1p$_{1/2}$ & 0.538 & 10.8&  9.2& 1.2  &21.2 \\ 
          &         &         &     & 2p$_{3/2}$ & 0.001 & 0.04& 0.03& 0.0  &0.07 \\ 
          &         &         &     & 2p$_{1/2}$ & 0.002 & 0.08& 0.06& 0.01  &0.14\\ 
\hline
\multicolumn{10}{|r|}{$\sigma_{p}^{Glauber}$=130 mb}\\
\hline
 $^{13}$O & $3/2^-$ & g.s.  & $1^+$ & 1p$_{3/2}$ & 0.086 & 3.4 & 3.4 & 1.03&7.83 \\
          &         &         &     & 1p$_{1/2}$ & 0.537 & 19.3& 19.1& 5.84&44.24 \\ 
\hline
\multicolumn{10}{|r|}{$\sigma_{p}^{Glauber}$=56 mb}\\
\hline
 $^{15}$O & $1/2^-$ & g.s.  & $1^+$ & 1p$_{3/2}$ & 0.372 & 6.5 & 5.5 & 0.65&12.65 \\
          &         &         &     & 1p$_{1/2}$ & 0.606 & 9.7 & 8.1 & 0.98&30.45 \label{tabre}\\ 
\hline
\multicolumn{10}{|r|}{$\sigma_{p}^{Glauber}$=46 mb}  \\
\hline
\hline
\end{tabular}
\end{center}

\end{table}

\end{document}